# Accurate Cathode Properties of LiNiO$_2$, LiCoO$_2$, and LiMnO$_2$ Using the SCAN Meta-GGA Density Functional

Arup Chakraborty, Mudit Dixit, Dan T. Major*

*Department of Chemistry, Bar-Ilan University, Ramat-Gan 52900, Israel*

Layered lithium intercalating transition metal (TM) oxides are promising cathode materials for Li-ion batteries. Here we scrutinize the recently developed strongly constrained and appropriately normed (SCAN) density functional method to study structural, magnetic and electrochemical properties of prototype cathode materials LiNiO$_2$, LiCoO$_2$, and LiMnO$_2$ at different Li-intercalation limits. We show that SCAN outperforms earlier popular functional combinations, providing results in considerably better agreement with experiment without use of Hubbard parameters, and dispersion corrections are found to have a small effect. In particular, SCAN provides a good description of the electronic structures, electron densities, band gaps, predicted cell parameters, and voltage profiles.

## 1. Introduction:

Renewable energy has emerged as one of the most important areas of research in recent years, due to growing concerns related to pollution, sustainability, and geopolitics. Leading green energy candidates today include solar cells, fuel cells, supercapacitors, and rechargeable batteries.[1-10] These technologies are in increasing demand, due to a global surge in energy consumption, widening dependence on electronic gadgets, and the desire to replace combustion engines in vehicles with more benign environmental alternatives. In these regards, Li-ion batteries (LIBs) have earned their place as one of the most promising technologies in the renewable energy field, and are currently

*E-mail: majort@biu.ac.il

employed in a variety of applications, ranging from cell phones and lap-tops to electrical vehicles.[1-10]

LIBs are multi-component devices and each component comes with scientific and technological challenges.[6] The performance in LIBs, such as energy density and capacity retention, may be greatly enhanced by improving the cathode.[3,6,11] Various types of cathode materials exist, including layered, spinel, olivine, and tavorite structures.[3,9,12-14] In particular, layered materials have surfaced as a very promising family of materials.[15,16] $LiCoO_2$ (LCO) was first introduced in 1980, and was the first layered transition metal oxide to be successfully incorporated into commercial rechargeable LIBs.[17] Today, LCO and other layered transition metal oxide cathodes are still widely used in portable electronic devices as positive electrodes. It has long been realized that mixing different transition metals (TM) in layered materials can be beneficial.[18,19] For instance, in Ni-Co-Mn layered materials, Ni provides favorable capacity, Co kinetics, and Mn stability.[15,16] To improve the specific capacity of cathodes, which is important for electrical vehicle applications, increasing the Ni-amount is a promising strategy.[20] The current state-of-the-art for these materials stands at capacities close to 300 $mAh \cdot g^{-1}$ for so-called Ni-rich materials.[15,16,21] However, Ni-rich layered cathode materials suffer from serious capacity fading during electrochemical cycling due to Ni-ion migrations and oxygen release.[21-24]

An important tool in studying cathode materials is density functional theory (DFT).[9,11,25-27] Using DFT, one can predict the structure, energetics, magnetism, electrochemical properties, and degradation mechanisms.[26] Early work in this area employed the local density approximation (LDA),[28-31] although the generalized gradient approximation (GGA)[32,33] was quickly adopted as the gold-standard in this area.[34] In particular, the Perdew-Burke-Ernzerhof (PBE) functional[35] has become the functional of choice.[34] In both LDA and GGA approaches, it is customary to include the so-called U Hubbard parameter for strongly correlated systems, which corrects for some of the electron self-interaction in DFT, by localizing the electrons.[36,37] Although this is a widely used approach, it does have numerous disadvantages. A trivial disadvantage is that the correct parameter must be derived, typically on a per element basis. A subtler disadvantage is that this parameter will not be identical for different oxidation states of a given element.[38] In the case of electrochemical application, this can be a problem as the oxidation states of transition metals change during Li-ion intercalation.[14,21,39] Additionally, use of a U parameter can significantly perturb the electronic structure of materials.[40] Hence, a U-free density functional approach, that better accounts for self-interaction, would be of great use. Hence, a more appropriate functional for electrochemical applications is sought.[41]

Recently, a new functional, dubbed strongly constrained and appropriately normed (SCAN),[42,43] was suggested. This functional, belongs to the non-empirical meta-GGA family of functionals, which includes the gradient of the kinetic energy density. In general, non-empirical functionals should satisfy a number of exact constraints. SCAN is the first semilocal meta-GGA exchange-correlation functional that satisfies 17 known possible exact constraints.[42] Furthermore, this functional accurately captures intermediate range weak interactions in non-bonded systems and rare-gas atoms due to appropriate norming.[42]

Peng and Perdew recently reported excellent binding energies and structural parameters for layered chalcogenides using the SCAN functional and an additional nonlocal correlation functional.[44] It has been also shown that the SCAN functional is a better choice for energetics and structural parameter of binary oxides than other functionals, such as PBE, LDA+U, and PBEsol.[45] Further, the SCAN functional has been applied to conventional perovskite ferroelectrics, like $BaTiO_3$, $CdTiO_3$ and $PbTiO_3$, showing improved prediction of structural, electric and energetic properties for these materials.[46,47] It was shown in a recent study that random phase approximation (RPA) and SCAN give good agreement for materials like hybrid perovskite.[48] The use of SCAN also provides a good description of the instability of the demagnesiated battery material $Cr_2O_4$.[49] There are also studies that suggest good performance of meta-GGA functionals like SCAN for energetics and electrochemical properties of Ni, Mn and Ti oxides.[50-52]

In the current work, we apply the SCAN functional to lithium intercalating layered transition metal oxide cathode materials to investigate the electronic, thermodynamic and electrochemical properties of three basic and widely used LIB cathode materials, namely, $LiNiO_2$ (LNO), LCO, and $LiMnO_2$ (LMO). We demonstrate that the SCAN functional largely outperforms both the GGA and GGA+U methods, offering better agreement with experimental data for most studied properties. Importantly, we show that the SCAN functional alleviates the need for the notorious U parameter in these basic LIB cathode materials.

## 2. Computational Details:

All DFT electronic structure calculations were performed using the plane wave based Vienna ab-initio Simulation Package (VASP).[53,54] All calculations employed projector augmented wave (PAW)[55] potentials for all elements. An energy cut-off of 520 eV was imposed for the plane wave basis. We considered an R-3m rhombohedral unit cell for LNO and LCO, and a Pmmn orthorhombic unit cell for LMO in our calculations (See Fig. 1 (a), 1 (b) and 1 (c), respectively). In LNO, LCO and LMO, the supercells were constructed from the primitive cell by doubling the system along the *a* direction. A Gamma-centered 4×8×2 *k*-mesh grid was used for the LCO and LNO supercells and a 6×4×3 *k*-mesh grid was used for LMO. Geometry and cell parameter optimizations were performed using the conjugate-gradient method and the convergence criteria was set to 0.01 eV/Å. We employed the PBE functional[35] (PBE) functional with and without onsite Coulomb interaction, U. The effective U parameters were 5.96 eV, 3.0 eV, and 5.10 eV for Ni, Co, and Mn, respectively,[14] and Dudarev's method[56] was used in GGA+U. Further, we employed the recently developed SCAN[42,43] exchange-correlation functional. Dispersion corrections were included using Grimme's DFT-D3 method.[57] Overall, we considered the following functional combinations: PBE, PBE+U, PBE+U+D3, SCAN and SCAN+D3.

## 3. Results and discussion:

In the following, we present the structural details, band-gap, magnetic and electronic structure, formation energy, and electro-chemical properties for LNO, LCO and LMO.

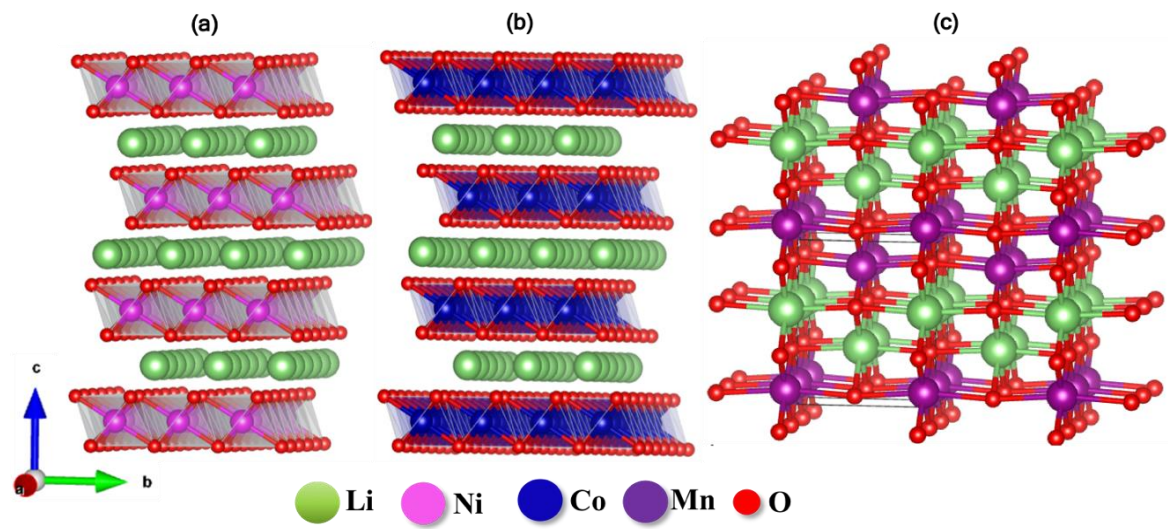

Figure 1. Structure of (a) LNO and (b) LCO in the hexagonal R-3m space group and (c) LMO in the Pmmn space group.

**3.1. Structural Parameters**:

The electrochemical properties of electrode materials depend significantly on the changes in structural parameters during cycling. We analyzed the changes in structural parameters of LNO, LCO and LMO at different delithiation levels ($x$=1.0, 0.5 and 0.0). From Fig. 2 we observe that the in-Li-plane $a$ parameter changes only slightly during delithiation, whereas the orthogonal vector, $c$, changes considerably, in line with previous studies on layered cathode materials.[14,21,24,39] Specifically, the $a$ parameter decreases slightly on delithiation for LNO and LCO, while it increases for LMO. This effect may be attributed to differences in the structure of LMO (orthorhombic), and LNO and LCO (rhombohedral). For all studied systems, the DFT methods give similar changes for the $a$ lattice vector on delithiation. The $a$ parameter represents the in-layer distance between two transition metals (TMs) in LNO and LCO, and decreases with delithiation since the ionic radius of the TMs decreases with increasing oxidation state of the TMs. We further observe a systematic monotonous increase in the $c$ lattice vector for LNO and LCO using the PBE and PBE+U methods. Interestingly, on including dispersion correction to PBE and PBE+U, the $c$ vector shows the expected and characteristic dip in the fully delithiated limit.[14,58,59] The $c$ parameter reprecents the interlayer distance between two TM layers in LNO and LCO, and it initially increases with delithiation due to electrostatic repulsion between adjacent O-layers, while close to the fully delithiated limit there is a decrease in interlayer slab distance. Interestingly, using the SCAN functional, this correct behavior is observed both with and without dispersion correction, as this nonlocal functional captures short-range weak interactions. The described variations of $a$ and $c$ lattice parameters agree with recent reports.[14,60-62] The changes in volume with delithiation is a combined effect of *a* and *c* parameter, but the *c* parameter provides the dominating effect. Therefore, the stability of the structure, and hence the capacity of a LIB cell depends significantly on the *c* parameter.[61] A comparison of calculated (using different DFT functionals) and experimental lattice parameters for fully lithiated systems is presented in Table 1. All approaches studied here perform well againts the experiemntal data, with SCAN and PBE-D3+U giving the best agreement for the three systems discussed here. We note that the calculations are 0 K temperature calculations and do not include any thermal breathing of the lattice.

Table 1: Lattice parameters of LNO, LCO and LMO computed using different functionals and from experiment.

| System | Lattice parameters | PBE | PBE+U | PBE+U+D3 | SCAN | SCAN-D3 | Expt. |
|---|---|---|---|---|---|---|---|
| **LNO** | $a$ (Å) | 2.887 | 2.841 | 2.807 | 2.823 | 2.796 | 2.876[63] |
| | $c$ (Å) | 14.202 | 14.289 | 14.124 | 14.094 | 13.932 | 14.191[63] |
| | Volume (Å$^3$) | 103.255 | 101.715 | 98.210 | 99.200 | 96.195 | |
| **LCO** | $a$ (Å) | 2.854 | 2.8375 | 2.812 | 2.8075 | 2.7845 | 2.830[63] |
| | $c$ (Å) | 14.054 | 14.152 | 13.978 | 13.959 | 13.793 | 14.119[63] |
| | Volume (Å$^3$) | 99.150 | 98.695 | 95.720 | 95.305 | 92.630 | |
| **LMO** | $a$ (Å) | 2.796 | 2.860 | 2.841 | 2.793 | 2.779 | 2.806[64] |
| | $c$ (Å) | 5.722 | 5.830 | 5.784 | 5.706 | 5.663 | 5.750[64] |
| | Volume (Å$^3$) | 76.320 | 77.450 | 75.635 | 73.040 | 71.165 | 73.89[64] |

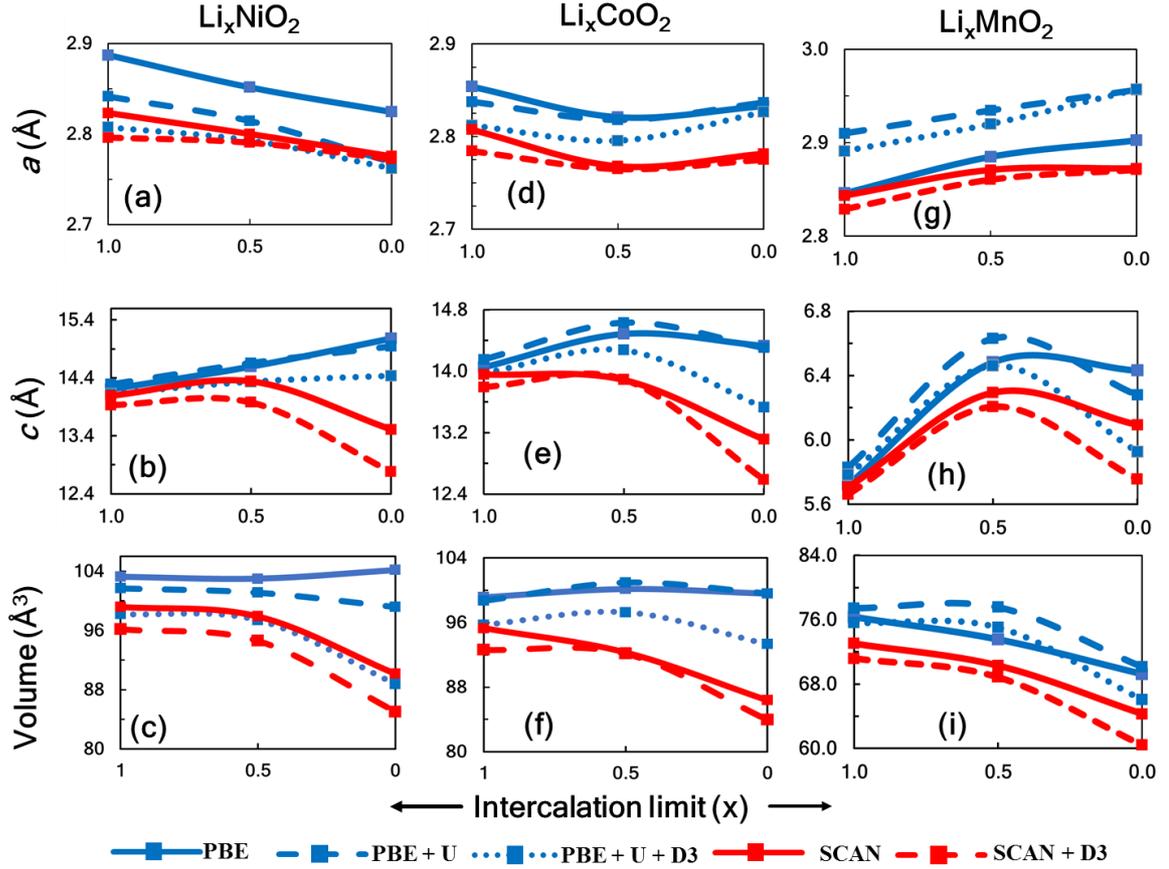

Figure 2. Variation in unit cell lattice parameters, *a*, *c* and volume, at different intercalation levels ($x$) for $Li_xNiO_2$, $Li_xCoO_2$, and $Li_xMnO_2$ using different functional combinations.

### 3.2. Band Gap:

To assess the electronic conductivity, we calculated the band gap for pristine LNO, LCO and LMO. LNO is predicted to be a half-metal for all functional combinations employed (Table 2), and this is in agreement with earlier computational data[36] and in reasonable agreement with experimental observations, which predict a very small band-gap.[63] For LCO, we obtain a band-gap of 1.08 eV using PBE, while adding a U parameter increases the band gap to 2.2 eV. The reported experimental band gap for LCO ranges from 2.1 to 2.7 eV. The SCAN functional gives a band-gap of 1.74 eV, showing better agreement with the experimental data than PBE, but not quite as good as PBE+U. The computed band gaps for LMO are 0.92 eV, 1.42 eV 1.19 eV using PBE, PBE+U and SCAN, respectively, which may be compared with the experimental band-gap of 1.64 eV.[65]

Table 2. Computed and experimental band gaps (eV) using different density functional combinations.

| System | PBE | PBE+U | PBE+U+D3 | SCAN | SCAN+D3 | Exp. |
|---|---|---|---|---|---|---|
| LiNiO$_2$ | HM$^a$ | HM$^a$ | HM$^a$ | HM$^a$ | HM$^a$ | 0.4[63] |
| LiCoO$_2$ | 1.08 | 2.22 | 2.26 | 1.74 | 1.96 | 2.1-2.7[66] |
| LiMnO$_2$ | 0.92 | 1.42 | 1.42 | 1.19 | 1.11 | 1.64[65] |

$^a$ HM = half-metal

### 3.3. Magnetic and electronic properties:

To examine the electronic and magnetic properties of LNO, we considered both parallel and antiparallel configurations for the Ni ions. Our total energy calculations show that the ferromagnetic (parallel), configuration is energetically favorable by 16 meV per formula unit (using PBE) and this result is in agreement with earlier computational studies.[38,63] In fully lithiated LNO, all the Ni ions are in a 3+ oxidation state and the low-spin (LS) electronic configuration for Ni$^{3+}$ is $t_{2g}^6$(|↑↓|↑↓|↑↓|) $e_g^1$(|↑ | |). Hence, the calculated local magnetic moment of Ni in LNO is expected to be ~1 $\mu_B$, and all methods give ca. 1 $\mu_B$ as shown in Table 3, in agreement with earlier reports.[36,63] The SCAN functional gives a magnetic moment slightly closer to unity than PBE. The local magnetic moments of Ni sites in partially and fully delithiated states are different because of the presence of Ni$^{4+}$, which has a $t_{2g}^6$(|↑↓|↑↓|↑↓|) $e_g^0$(| | |) configuration.

Table 3. Average local magnetic moment ($\mu_B$) of transition metals in pristine $LiNiO_2$, $LiCoO_2$, and $LiMnO_2$ using different functionals.

|     | PBE  | PBE+U | PBE+U+D3 | SCAN | SCAN+D3 |
|-----|------|-------|----------|------|---------|
| **LNO** | 0.76 | 1.12  | 1.04     | 0.86 | 0.84    |
| **LCO** | 0.00 | 0.00  | 0.00     | 0.00 | 0.00    |
| **LMO** | 3.49 | 3.90  | 3.89     | 3.60 | 3.58    |

We further analyzed the electronic structure of pristine LNO using the different functionals (Fig. 3(a), (b) and (c), respectively). We observe hybridization between Ni-d and O-p states and there is a finite density of states (DOS) at the Fermi level for all methods. Inspection of the DOS in Fig. 3(b) reveals that both up and down $t_{2g}$ spin channels are completely occupied, while the up channel of $e_g$ is partially occupied, indicating that $Ni^{3+}$ is in LS, as discussed above. Interestingly, using PBE and PBE+U the $t_{2g}$ and $e_g$ states are similarly occupied, whereas with SCAN $t_{2g}$ is largely occupied and $e_g$ is unoccupied.

In the experimental valence electron XPS for LNO it was observed that the band near the Fermi level (at ~ -1.4 eV) is predominantly composed of Ni-3d, while O-2p states were found at ~ -3.8 eV.[67] Here, we find that both PBE and SCAN predict dominant contribution of Ni-3d states near the Fermi level (-1 to -2 eV) for LNO, in agreement with experimental data,[67] while PBE+U predicts dominant contribution of O-2p states in this range. The deleterious effect of U on the relative d- and p-band positions in pristine layered oxide materials has been noted in our previous study on Ni-rich mixed transition metal oxides.[39] We underscore that the dominant contribution of Ni-3d states near the Fermi level, that is correctly predicted by SCAN, explains classical redox behavior of LiNiO$_2$. We further note that the electronic structure predicted by SCAN also suggests a slightly higher degree of hybridization than that of the PBE method.

The calculated local magnetic moment on Co sites in the fully lithiated LCO is 0.0, because the Co$^{3+}$ ions are in a low-spin state (LS) (t$_{2g}^6$(|↑↓|↑↓|↑↓|) e$_g^0$(| | |)), in agreement with earlier studies on LCO.[63] The electronic structure (DOS) is displayed in Fig. 3 (d), (e) and (f) for the PBE, PBE+U and SCAN functionals, respectively. From Fig. 3 (d), (e) and (f), we observe strong hybridization between Co-3d and O-2p. Further, we note that with the SCAN functional, the t$_{2g}$ band of Co is completely occupied in both spin channels, while e$_g$ is unoccupied, reflecting the LS state of Co, as explained above. The DOS is in agreement with XPS studies of LCO, which showed similar hybridization between Co-3d and O-p states.[66] Ensling et al. performed valence XPS of LCO and noted a dominant contribution of Co-3d states in the upper valence band region (~ -1-3 eV). They also suggested that a broad band near -2.5-7.5 eV has prevalent O-2p character, but with some Co 3d admixtures.[68] We note that the electronic structure obtained using both PBE and SCAN suggest dominant contribution of Co-3d states in the upper valence region, in agreement with experiments,[68] and suggest a classical redox behavior for Co in the fully intercalated limit. However, PBE+U suggest almost equal contribution of Co-3d and O-2p states near the Fermi level. We note that on increasing the U value, the contribution of the O-2p states near the Fermi increases. We further note that the occupied and unoccupied bands are more separated in PBE+U compared to PBE and it effectively increases the band gap of the system. However, using PBE+U the valence band was found to be pinned to the Fermi level unlike the results obtained with PBE, SCAN and the experimental valence XPS.[68]

For LMO, the antiferromagnetic (AFM) spin configuration is energetically favorable for all functional combinations investigated. Using PBE, the difference of total energy between AFM and ferromagnetic configurations per formula unit is 0.125 meV, while using SCAN it is 0.085 meV. The local magnetic moment of Mn (~ 3.6 $\mu_B$) implies that Mn is in a high spin state $t_{2g}^3$(|↑ |↑ |↑ |) $e_g^0$(|↑ |  |). In Fig. 3 (g), (h) and (i), we display the DOS for LMO in the AFM configuration, and we observe strong hybridization between Mn-d and O-p near the Fermi level using SCAN, in agreement with the experimental XPS data.[69] However, PBE+U suggests significantly greater contribution of O-2p states in the upper valence band region.

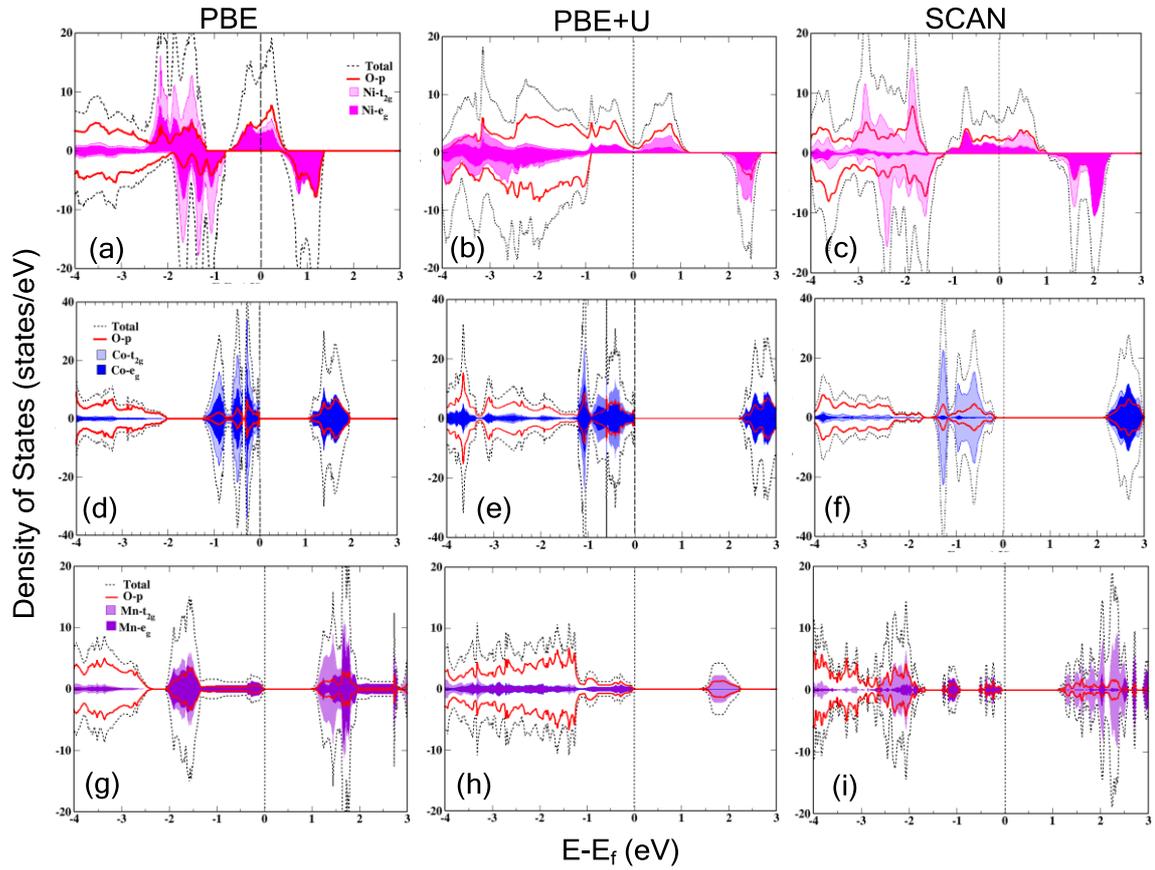

Figure 3. Density of States for LNO (a, b and c), LCO (d, e and f) and LMO (g, h and i) obtained with the PBE, PBE+U, and SCAN functionals. The systems were rotated to align the MO$_6$ octahedra with the global z-axis of the unit-cell to assign the $t_{2g}$ and $e_g$ states.

We also assessed the quality of the electron density of the different methods (Fig. 4). To this end we compared the electron density of PBE, PBE+U, and SCAN to the electron density from the hybrid density functional PBE0,[70] which has recently been shown to give excellent electron density relative to high-level ab-initio methods.[71,72] Visual inspection of the density difference figures clearly show that SCAN gives better agreement with PBE0 than the other functionals. The quantitative difference in density (*DD*) was estimated according to:

$$DD = \sqrt{\frac{\sum_{i=1}^{N}(\rho_i(XC)-\rho_i(PBE0))^2}{N}} \qquad (1)$$

Here, $\rho_i$ is the density at grid point *i*, *XC* is one of functionals PBE, PBE+U, or SCAN, while *N* is the number of grid points. The quantitative analysis (orange boxes in Fig. 4) confirms the impression from the visual inspection, that the electron density from SCAN is the most accurate among the functionals tested.

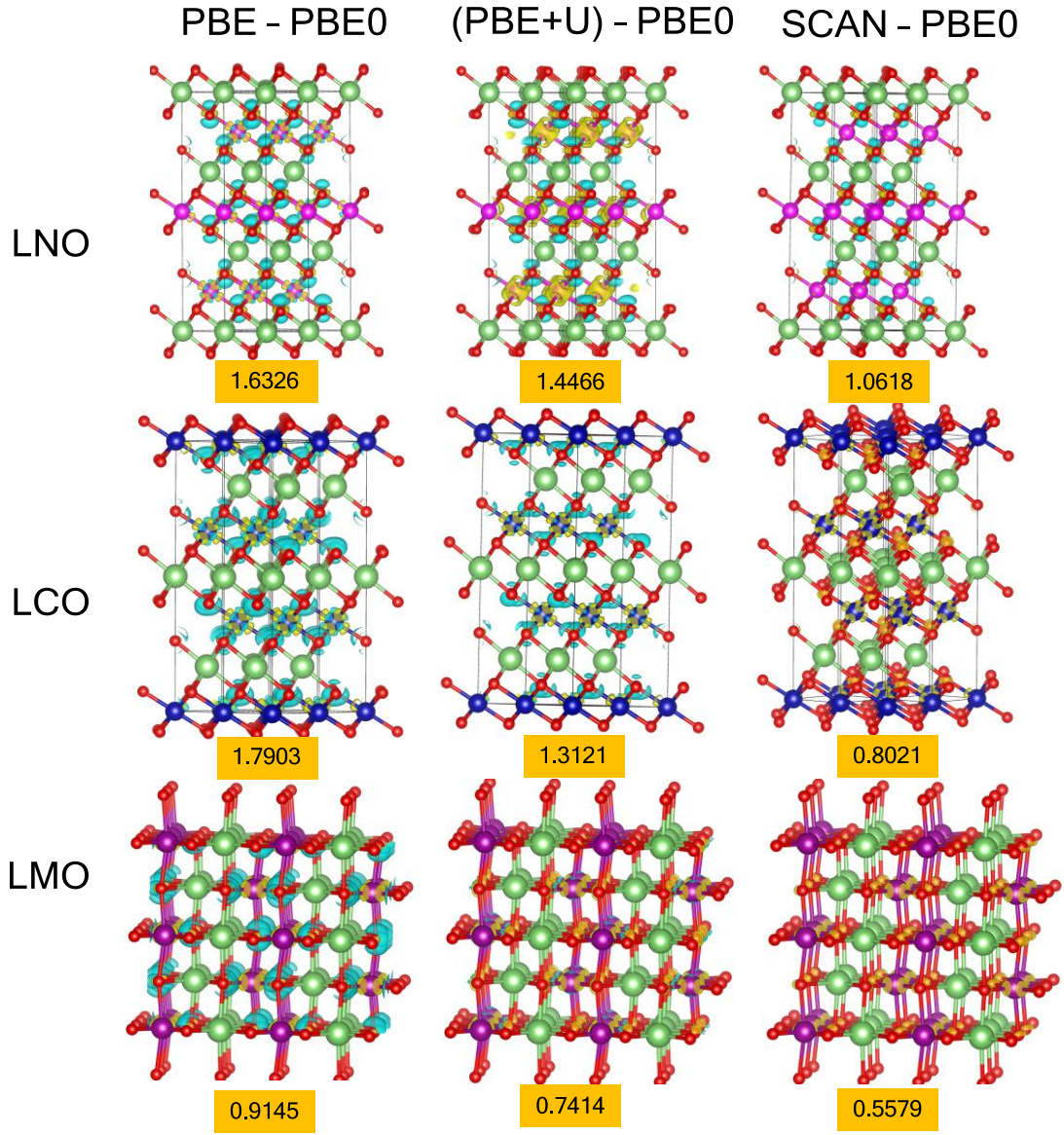

Figure 4. Density difference isosurfaces (blue/yellow colors) between the three target DFT methods PBE, PBE+U, and SCAN, and the density of the hybrid functional PBE0. The isodensity value was set to 0.003352 a.u. The quantitative difference is shown in orange boxes (a.u.).

### 3.4. Formation energy:

To understand the formation of solid solutions of lithiated and delithiated LNO, LCO and LMO, we calculated the formation energy per formula unit:[73]

$$FE = E(Li_xMO_2) - x \cdot E(LiMO_2) - (1-x) \cdot E(MO_2) \quad (2)$$

Here $x$ is the fractional amount of Li present in the system, E(Li$_x$MO$_2$) is the energy of the partially delithiated material, while E(LiMO$_2$) and E(MO$_2$) represent the energies of the pristine and fully delithiated structures, respectively. The calculated formation energy for different intercalation limits using different functionals is shown in Fig. 5 (a), 4(b) and 4(c) for LNO, LCO and LMO respectively. We observe that the formation energy of partially lithiated states is negative for all materials, and hence a solid solution is predicted. This preference is minute using SCAN, while more significant using PBE. In an earlier study using PBE, the FE for LCO is calculated to be ca. 0.22 eV/f.u.[73] Here we observe that the FE for partially lithiated (x=0.5) LCO using PBE and SCAN are calculated to be 0.21 and 0.16 eV/f.u. respectively. Similarly, our calculated value of 103 meV using PBE for partially lithiated (x=0.5) LNO is in good agreement with the value reported by Ceder and co-workers using the same functional (~130 meV/f.u.).[25] Using SCAN this value is somewhat lower. We note that all the applied functionals predict a solid solution behavior in agreement with experiments.[58,74,75]

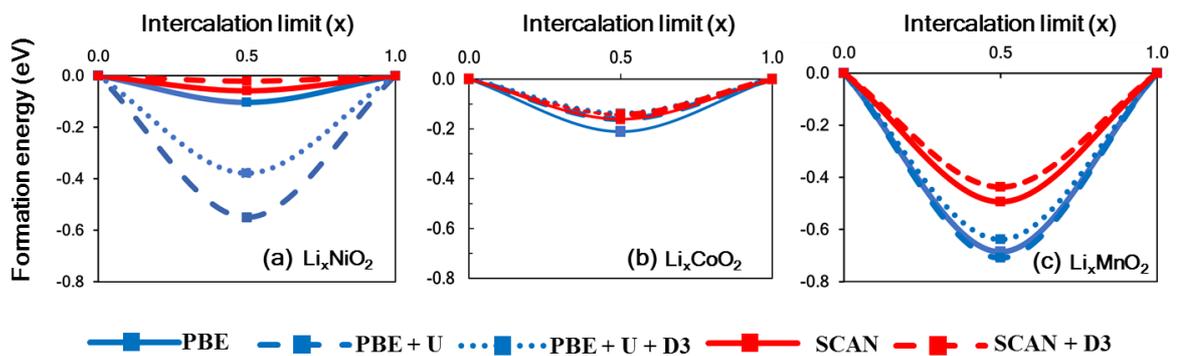

Figure 5. The formation energy for (a) Li$_x$NiO$_2$, (b) Li$_x$CoO$_2$, (c) Li$_x$MnO$_2$ at different intercalation limits ($x$) using different functional combinations.

**3.5. Intercalation voltage:**

We calculated the intercalation voltage using the following formula:[14,29]

$$V = -\frac{E(Li_{x+dx}MO_2) - E(Li_xMO_2)}{dx} + E(Li_{bcc}) \qquad (3)$$

Where $E(Li_{x+dx}MO_2)$ and $E(Li_xMO_2)$ represent the total energy per formula unit of the system before and after lithium deintercalation. $E(Li_{bcc})$ is the energy per formula unit of bulk Li. The calculated intercalation potentials are displayed in Fig. 6. For LNO, the experimental voltage profile for the fully lithiated to fully delithiated states varies from 3 V to ~4.3 V in LNO.[25,76] The calculated intercalation profile for LNO ranges from 2.8 to 3.2 V and 3.6 to 3.8 V using PBE and SCAN, respectively, while using PBE+U results in a significant overestimation. The observed intercalation profile for LCO ranges from 3.0 to 3.8 V and 4.1 to 4.7 V using PBE and SCAN, respectively, whereas the experimental profile varies from 3.6 V to 4.8 V.[77] The observed intercalation profile for LMO ranges from 2.4 to 3.2 V and 2.9 to 3.4 V using PBE and SCAN, respectively, while PBE+U is within the experimental range. Experimentally, the voltage ranges from 2.0 V to 4.6 V for LMO.[16,78] In conclusion, the intercalation voltage is underestimated in case of PBE, while using the SCAN functional the results are in better agreement with experiments.

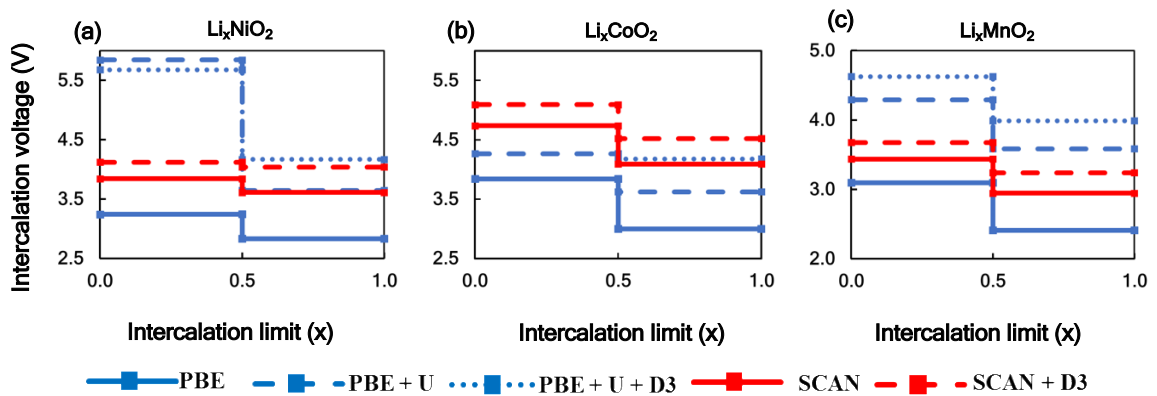

Figure 6. Intercalation potentials (V) for (a) $Li_xNiO_2$ (b) $Li_xCoO_2$ and (c) $Li_xMnO_2$.

## 4. Summary

We studied the structural details, band-gap, magnetic and electronic structure, formation energy, and intercalation profiles for LNO, LCO and LMO using standard GGA density functional approaches and the recently developed meta-GGA SCAN density functional. The computed data were compared with available experimental data, and we observed that the SCAN functional performs similarly or better for most properties studied. Importantly, the SCAN functional alleviates the need for the notorious U parameter, and the effect of dispersion correction is modest. In particular, SCAN gives good structural behavior during electrochemical cycling, and provides better electronic structure than PBE+U and better band-gaps than PBE. Finally, SCAN predicts voltage profiles in better agreement with experiment than PBE and PBE+U. In conclusion, SCAN is a versatile functional that provides good all-round performance for all relevant electrochemical properties benchmarked in this study.

## Acknowledgements

This work was partially supported by the Israel Science Foundation (ISF) in the framework of the INREP project.